y\def\sqr#1#2{{\vcenter{\vbox{\hrule height.#2pt \hbox{\vrule
width.#2pt height#1pt \kern#1pt \vrule width.#2pt} \hrule height.#2pt}}}}

\documentstyle[12pt]{article}
\input epsf
\begin{document}
\begin{titlepage}
\title{Quantum-to-classical Transition of Cosmological Perturbations for
Non-vacuum Initial States}
\author{J. Lesgourgues$^{1}$, David Polarski$^{1,2}$ and A. A. Starobinsky$^a$\\
\hfill \\
$^1$~{\it Laboratoire de Math\'ematiques et de Physique Th\'eorique, EP93
CNRS}\\
{\it Universit\'e de Tours, Parc de Grandmont, F-37200 Tours (France)}\\
\hfill\\
$^2$~{\it D\'epartement d'Astrophysique Relativiste et de Cosmologie},\\
{\it Observatoire de Paris-Meudon, 92195 Meudon cedex (France)}\\
\hfill \\
$^a$~{\it Landau Institute for Theoretical Physics}, \\
{\it Kosygina St. 2, Moscow 117334 (Russia)}}

\date{\today}
\maketitle

\begin{abstract}
Transition from quantum to semiclassical behaviour and loss of quantum
coherence for inhomogeneous perturbations generated from a non-vacuum initial
state in the early Universe is considered in the Heisenberg and the
Schr\"odinger representations, as well as using the Wigner function. We show
explicitly that these three approaches lead to the same prediction in the
limit of large squeezing (i.e. when the squeezing parameter $|r_k|\to \infty$):
each two-modes quantum state $({\bf k},~-{\bf k})$ of these perturbations is
equivalent to a classical perturbation that has a stochastic amplitude,
obeying a non-gaussian statistics which depends on the initial state,
and that belongs to
the quasi-isotropic mode (i.e. it possesses a fixed phase). The Wigner
function is not everywhere positive for any finite $r_k$,
hence its interpretation as a classical distribution function
in phase space is impossible without some coarse graining procedure.
However, this does not affect the transition to semiclassical
behaviour since the Wigner function becomes concentrated near a
classical trajectory in phase space when $|r_k|\to \infty$ even without
coarse graining. Deviations of the statistics of the perturbations in real 
space from a Gaussian one lie below the cosmic variance level for the 
$N$-particles initial states with $N=N(|{\bf k}|)$ but may be observable for 
other initial states without statistical isotropy or with correlations between 
different ${\bf k}$ modes. As a way to look for this effect, it is proposed to 
measure the kurtosis of the angular fluctuations of the cosmic microwave 
background temperature.
\end{abstract}

PACS Numbers: 04.62.+v, 98.80.Cq, 98.80.Hw
\end{titlepage}

\section{Introduction}
The undeniable success of the theory of the generation of cosmological
perturbations from quantum vacuum fluctuations during an inflationary era
in the early Universe was its ability to correctly predict in advance three
main properties of the recently discovered (non-dipole) angular fluctuations
$\frac{\Delta T(\theta, \phi)}{T}$ of the cosmic microwave background (CMB)
temperature. These predictions are:
a) the multipole power spectrum of the fluctuations that is
approximately flat for low multipoles
($C_l\propto \frac{1}{l(l+1)}~\rm{for}~ 2\leq l\leq 30$);
b) the right location and amplitude of the first acoustic (or ``Doppler'')
peak of $\frac{\Delta T}{T} \rm$ at $l=200-250$;
c) the Gaussian statistics of the fluctuations (see, e.g., the 4-year COBE
results for the spectrum~\cite{cobe1}
and statistics~\cite{cobe2} of the fluctuations with $l\leq 20$, and the
results of the Saskatoon-95 experiment for $l=80-350$~\cite{sas96} and
the CAT experiment for $l=400-600$~\cite{cat96}). Of course, this does not
exclude alternative theories of the early Universe (e.g., topological
defects) provided they are able to produce approximately the same
observational predictions. However, it is clear that we can and even must
seriously investigate quantum-gravitational effects in the early Universe
because of the actual observability of their consequences at present time.

At this point, one of the most important and interesting questions of this
theory arises. The initial perturbations generated during an inflationary
stage were quantum, relativistic and gravitational since their
dimensionless amplitude contains all three fundamental constants: the Planck
constant $\hbar$, the light velocity $c$ and the Newtonian gravitational
constant $G$ (below we will take $c=\hbar=1$ unless otherwise stated). The
fluctuations $\frac{\Delta T}{T}$ measured today, on the other hand, are
certainly classical (though stochastic). Thus, some mechanism describing the
quantum-to-classical transition of the perturbations is required. What was
explicitly or implicitly used in the pioneering papers where the spectrum of
tensor perturbations (gravitational waves)~\cite{staro79} and scalar
(adiabatic) perturbations~\cite{hawking82} generated from vacuum fluctuations
was first correctly derived, was a kind of equivalence between the strongly
squeezed vacuum fluctuations and classical perturbations with a stochastic
Gaussian amplitude and a fixed temporal phase in the Heisenberg
representation (see~\cite{cqg96} for a detailed discussion and explanation
of this equivalence). The same applies to the Schr\"odinger representation,
see~\cite{hal85} where the quantum cosmological approach based on the
Wheeler-de Witt equation was used, it reduces to the Schr\"odinger
representation for each mode after separation of variables and introduction
of the quasi-classical time. Finally, this equivalence can be seen as well
using the Wigner function formalism~\cite{cqg96,guth85}. An essential
property of this approach is that this type of transition from a quantum to a
classical description of the perturbations, including the complete loss of
quantum coherence, does not require any actual consideration of very small
interactions between a given mode of perturbations and other degrees of
freedom (``environment'') in the Universe. Though these interactions do of
course exist and do lead to the transformation of an initial pure quantum
state into a mixed state, their actual structure is unimportant for
observable quantities as far as the exponentially small decaying mode of
perturbations may be neglected (note however that it must be taken into
account if one wants to calculate the entropy of cosmological
perturbations~\cite{david96}). This is precisely why this transition was
called ``decoherence without decoherence'' in~\cite{cqg96}. Furthermore,
though the quantum coherence disappears, not all coherence is lost: indeed,
the equivalent classical stochastic perturbation remains partly correlated
(coherent) at the classical level in the sense that it has a fixed temporal
phase corresponding to the fact that it belongs to the special
quasi-isotropic mode of perturbations. Remarkably, this coherence has
observable consequences: it results, in particular, in the appearance of
multiple peaks (called acoustic, or Doppler peaks)
in the multipole power spectrum of $\frac{\Delta T}{T}$ with periodic spacing
$\Delta l=\pi \frac{\eta_0-\eta_{rec}}{\eta_0 v_s}$ in the case of adiabatic
perturbations.
Here $v_s$ is the sound velocity in the
relativistic plasma consisting of coupled baryons and photons just before
recombination, $\eta_o$ resp. $\eta_{rec}$ are the conformal times today resp.
at recombination.
A similar effect (though of significantly smaller amplitude)
takes place for CMB angular temperature fluctuations produced by primordial
gravitational waves background~\cite{cqg96} ($v_s=1$ in this case).

This approach however, especially that one based on the Wigner function
formalism, was strongly criticised in~\cite{anderson90} on the ground that
it might be applicable to vacuum initial conditions only and not to more
general ones. Hence, we adress in this paper the problem of the quantum to
classical transition of quantum cosmological perturbations in the case of
non-vacuum initial states. By an "initial state" of a given Fourier mode
$\bf k$ of a quantum field we mean its state in the WKB regime deep inside
the Hubble radius at the inflationary stage, long before the Hubble radius
crossing, i.e. when $k\gg aH$ where $k=|\bf k|$, $a(t)$ is the scale
factor of the Friedmann-Robertson-Walker (FRW) isotropic cosmological model,
$H\equiv \dot a/a$ is the Hubble parameter 
and the dot means a time derivative with respect to the cosmological time $t$.

Of course, a non-vacuum initial state contradicts the whole spirit of the
maximally symmetric initial state of the Universe which
lies at the heart of the inflationary scenario. Thus, our ``wish'' would
rather be to show that such an assumption is in conflict with observational
data. Surprisingly, as we will see below, this turns out to be more delicate.
Non-vacuum initial states are typically not self-consistent
in the simplest variants of the inflationary scenario since they
lead to a large energy density of inflaton field quanta which is not of the 
type of a cosmological term and it would contradict the assumption of 
inflation at time $\eta_0$ especially if we push $\eta_0$ to $-\infty$.
However, they could
appear in more complicated, double inflationary models for perturbation modes
which cross the Hubble radius three or more times during inflation
if we additionally assume (without justification) that these modes
immediately decohere (e.g., in the particle number basis) once they reenter
inside the Hubble radius. Though the latter process is hardly physically
viable, still it cannot be completely excluded. Furthermore, it is an
important question of principle to study whether the possibility of a
classical description of highly squeezed perturbations is generic or
whether it is restricted to vacuum initial states only.

Particle creation from non-vacuum initial states was considered in many
papers; in the cosmological case -  beginning with the pioneering paper by
Parker~\cite{par69}, see the recent review~\cite{hu94} for many references.
However, the primary quantities studied in these papers were the number
of created particles and the average value of the energy-momentum
tensor of a quantum field on a FRW background. In contrast, we are mainly
interested in the field fluctuations themselves since these are the quantities
(and not the number of particles) that are directly observable
in the case of cosmological perturbations. This refers both to scalar
metric, or density, perturbations and to tensor metric perturbations or 
gravitational waves. Naturally, the resulting quantum-to-classical
equivalence that, as we argued in~\cite{cqg96}, is the reason for the
possibility of a classical description of the perturbations in the present 
Universe takes place for these quantites, too.

Thus, the results of paper~\cite{cqg96} are generalized to the case of
non-vacuum initial conditions. In order not to introduce any preferred 
direction in space, we will consider only reflection symmetric 
$({\bf r} \to -{\bf r},~{\bf k}\to -{\bf k})$ quantum states.  
We show that a wide class of initial
states become quasiclassical
in the limit of high squeezing. This is shown in section 2 using the
Schr\"odinger representation, and in section 3 using the Heisenberg
representation. In section 4 we derive conditions necessary for the 
semiclassical behaviour of quantum scalar (adiabatic) perturbations
and gravitational waves on a FRW background when expressed directly in terms
of {\it rms} values of the perturbations. The Wigner function approach to 
this problem is studied in section 5. The specific, but important case of a 
thermal initial state is considered in section 6.
Finally, section 7 contains conclusions and a discussion of a possible
observational confirmation or rejection of a non-vacuum initial state.

\section{Schr\"{o}dinger representation}

In this section, we are going to compute the wave function of different
non-vacuum
states in the Schr\"odinger representation. Let us first review the basics
about the dynamics of the system that we consider here, namely
that of a real massless scalar field on a flat FRW
universe. Actually, the field can be massive and all the arguments presented
here will go through.
The space-time metric has the form
\begin{equation}
ds^2=dt^2-a^2(t)\delta_{ij}dx^i dx^j, \qquad\ i,j=1,2,3.
\end{equation}

Let us remind how the dynamics of this system will lead to the appearance
of squeezed states.
We first write down the Hamiltonian $H$ in terms of the field 
$y\equiv a\varphi$ and the conformal time $\eta =
\int {dt \over a(t)}$. The following result is then obtained
\begin{eqnarray}
H &=& \int d^3{\bf x}~{\cal H}(y, p, \partial_i y, t)\nonumber\\
&=& {1\over 2}\int d^3{\bf k} \lbrack p({\bf k})p^{\dagger}({\bf k})+k^2
y({\bf k})y^{\dagger}({\bf k})+{{a'}\over a} \left(y({\bf k})p^{\dagger}
({\bf k})+p({\bf k})y^{\dagger}({\bf k})\right) \rbrack
\label{H}
\end{eqnarray}
where
\begin{equation}
p\equiv{{\partial {\cal L}(y,y')}\over {\partial y'}}=y'-\frac{a'}{a}y
\end{equation}
and a prime stands for derivation with respect to conformal time.
Here the following Fourier transform convention is used:
$\Phi({\bf k})\equiv (2\pi)^{-3/2}\int \Phi({\bf r})e^{-i{\bf kr}}
d^{3}{\bf r}$ for functions as well as for operators.
Due to reality of the field $y$, we have that $y({\bf k})=y^{*}
(-{\bf k})$, resp. $y^{\dag}(-{\bf k})$ for operators. Therefore,
any classical field configuration is completely specified by
giving the Fourier transforms in half Fourier space.
Though generally this doesn't have to be the case when the fields are 
quantized, we will deal here with states which are invariant under the 
reflection ${\bf k}\to -{\bf k}$ as mentioned above.
The system can be described with the help of the complex valued field
modes $f_k(\eta)$ with $\Re f_k\equiv f_{k1}$ and $\Im f_k\equiv f_{k2}$,
$f_k(\eta_0)=1/\sqrt{2k}$ (we adopt a similar notation for all quantities),
and momentum modes $g_k(\eta)$,~$g_k(\eta_0)=\sqrt{\frac{k}{2}}$, typically
used in the Heisenberg representation
\begin{eqnarray}
y({\bf k},\eta)
&\equiv& f_k(\eta)~a({\bf k},\eta_0)+f_k^*(\eta)~
a^{\dag}(-{\bf k},\eta_0)~,\label{f}\\
p({\bf k},\eta)
&\equiv& -i\left(g_k(\eta)~a({\bf k},\eta_0)
-g_k^*(\eta)~a^{\dag}(-{\bf k},\eta_0)\right)~.\label{p}
\end{eqnarray}
In terms of $f_k~{\rm and}~g_k$, the important quantities
$\gamma_k~{\rm and}~F(k)$ are given by the following expressions
\begin{equation}
\gamma_k=\frac{g^*}{f^*}=\frac{1}{2|f_k|^2}(1-i2F(k))~,
\qquad\ F(k)=f_{k1}g_{k2}-f_{k2}g_{k1}~. \label{defF}
\end{equation}
The functions $f_k$ satisfy the equation
\begin{equation}
f_k'' + \left(k^2-{a''\over a}\right)f_k=0~, \label{weq}
\end{equation}
and
\begin{equation}
g_k = i\left(f_k'-{a'\over a}f_k\right)~.
\end{equation}

The couple of functions $f_k,~g_k$ represent only three independent real
functions as they satisfy a Wronskian constraint.
The dynamics of the system can also be
conveniently parametrized by the following three functions: the rotation angle
$\theta_k$, the squeeze angle $\phi_k$ and the squeeze parameter $r_k$. When
$|r_k|\to \infty$, the system is said to be highly squeezed. In that case also
$|F_k|\to \infty$.
As a result of the time evolution of our system, we have
\begin{eqnarray}
S~a^{\dagger}(\pm {\bf k},\eta_0)~S^{-1}
& = &
g_k(\eta)~y(\mp {\bf k},\eta_0) - i~f_k(\eta)~p(\mp {\bf k},\eta_0)~,
\nonumber \\
S~a(\pm {\bf k},\eta_0)~S^{-1}
& = &
g_k^*(\eta)~y(\pm {\bf k},\eta_0) + i~f_k^*(\eta)~p(\pm {\bf k},\eta_0)
\end{eqnarray}
where $S$ is the unitary evolution operator from time $\eta_0$\ to time
$\eta$.
When working in the Schr\"odinger (complex) coordinate representation, we will
often adopt the following compact notation
$y({\bf k},\eta_0) \equiv y_0,~
p({\bf k},\eta_0)\equiv p_0 = -i\frac{\partial}{\partial y^*_0},~
y(-{\bf k},\eta_0) \equiv y_0^*,~
p({-\bf k},\eta_0)\equiv p_0^* =-i \frac{\partial}{\partial y_0}$
where $y_0$\ and $y_0^*$\ are considered as independent variables.
\par
As a result of the coupling with the gravitational
field, which plays here the role of an external classical field, the state
$|0,~\eta \rangle_S$ which is the vacuum state of the field at some
given initial time $\eta_0,~a({\bf k},\eta_0)~|0,~\eta_0\rangle=0~
\forall {\bf k}$, will no longer be the vacuum at
later times $\eta$. Indeed field quanta are produced in pairs with
opposite momenta, we get a two-modes squeezed state. Though the state at late
time is no longer annihilated by the annihilation operators $a({\bf k})$, it
is still annihilated by the following time-dependent operator
\begin{equation}
\Bigl \lbrace y({\bf k},\eta_0)+i\gamma_k^{-1}(\eta)~p({\bf k},\eta_0)
\Bigr \rbrace|0,\eta\rangle_S=0~.\label{eq1}
\end{equation}
In the infinite volume case, equation (\ref{eq1}) translates into the
following functional differential equation in the Schr\"odinger (complex)
coordinate representation
\begin{equation}
\Bigl \lbrace y({\bf k},\eta_0)+\gamma_k^{-1}(\eta)\frac{\delta}
{\delta y({-\bf k},\eta_0)}
\Bigr \rbrace|0,\eta\rangle_S=0~.\label{eq2}
\end{equation}
Therefore the state at late time is still Gaussian, its wave functional in
the infinite volume case (the quantum variables are the functions
$y({\bf k})$ of the continuous variable $\bf k$) being given by
\begin{equation}
\Psi_0[y({\bf k},\eta_0),y(-{\bf k},\eta_0)]=
{\cal N}_k \exp \left(-\frac{1}{2}\int d^3{\bf k}~\gamma_k~~
y({\bf k},\eta_0)~y(-{\bf k},\eta_0) \right)~.
\end{equation}

For our purposes however, it is enough to consider our system enclosed in a
finite volume and we then get as wave function of our system an (infinite)
product of functions
\begin{equation}
\frac{1}{\sqrt{\pi}|f_k|}\exp \left(-\frac{|y({\bf k},\eta_0)|^2}{2|f_k|^2}
\lbrace 1-i2F(k)\rbrace \right)\label{Psi}
\end{equation}
for each pair ${\bf k},-{\bf k}$. The commutators and correlation functions
of Fourier transform operators are then c-number functions and not
distributions. The time dependence of $\Psi$ is through $f_k,~F(k),~{\rm and}~
N_k$. We note the persisting symmetry under reflections in $\bf k$-space, it
is present in the initial state and the evolution doesn't spoil it.

As a first generalization of the vacuum state at time $\eta_0$, we consider a
two-modes coherent state $|\alpha\beta,~\eta_0\rangle_S$ satisfying
\begin{eqnarray}
a({\bf k},\eta_0)~|\alpha\beta,~\eta_0 \rangle_S & = &
\Bigl\lbrace \sqrt{\frac{k}{2}} y({\bf k},\eta_0)
+\frac{i}{\sqrt{2k}} p({\bf k},\eta_0) \Bigr\rbrace~|\alpha\beta,~\eta_0
\rangle_S \nonumber\\
& = & \alpha~|\alpha\beta,~\eta_0 \rangle_S
= \Bigl\lbrace \sqrt{\frac{k}{2}}~\langle y\rangle_0 +\frac{i}{\sqrt{2k}}~
\langle p\rangle_0 \Bigr\rbrace~|\alpha\beta,~\eta_0 \rangle_S
\label{cohinst}
\end{eqnarray}
where $\langle y\rangle_0$ and $\langle y\rangle_0$ are the average values
of $y({\bf k},\eta_0)$ and $p({\bf k},\eta_0)$ at time $\eta_0$, and it
satisfies also
\begin{eqnarray}
a(-{\bf k},\eta_0)~|\alpha\beta,~\eta_0 \rangle_S
& = &
\Bigl\lbrace \sqrt{\frac{k}{2}} y(-{\bf k},\eta_0)
+\frac{i}{\sqrt{2k}} p(-{\bf k},\eta_0) \Bigr\rbrace~|\alpha\beta,~\eta_0
\rangle_S\nonumber\\
& = & \beta~|\alpha\beta,~\eta_0 \rangle_S=
\Bigl\lbrace \sqrt{\frac{k}{2}} \langle y\rangle_0 ^* +\frac{i}{\sqrt{2k}}
\langle p\rangle_0^*\Bigr\rbrace~|\alpha\beta,~\eta_0 \rangle_S~.
\end{eqnarray}
It is clear that the time-evolved state $|\alpha\beta,~\eta \rangle_S$
satisfies the following equations at any later time $\eta$
\begin{eqnarray}
S~a({\bf k},\eta_0)~S^{-1}~|\alpha\beta,~\eta \rangle_S
& = & \alpha~|\alpha\beta,~\eta \rangle_S =
(~g_k^*(\eta)~\langle y \rangle + i~f_k^*(\eta)~\langle p \rangle~)
~|\alpha\beta,~\eta \rangle_S\nonumber \\
S~a(-{\bf k},~\eta_0)~S^{-1} |\alpha\beta,~\eta \rangle_S
& = & \beta~|\alpha\beta,~\eta \rangle_S =
(~g_k^*(\eta)~\langle y \rangle^* + i~f_k^*(\eta)~\langle p \rangle^*~)
~|\alpha\beta,~\eta \rangle_S \label{coheq}
\end{eqnarray}
where $\langle y \rangle,\langle p \rangle$ are average values at time $\eta$.
It follows from (\ref{cohinst},\ref{coheq}) that $\langle y \rangle$ is the 
solution
of eq. (\ref{weq}) with the initial conditions $y(\eta_0)=\langle y\rangle_0~,
~p(\eta_0)=\langle p\rangle_0$, and $\langle p\rangle = \langle y \rangle'
-{a'\over a}\langle y \rangle$. Eqs. (\ref{coheq}) are easily solved in the 
Schr\"odinger representation resulting in the following 
wave function $\Psi_{\alpha\beta}(y_0,y_0^*,\eta)$ of
the two-modes coherent state: 
\begin{eqnarray}
\Psi_{\alpha \beta}(y_0,~\eta) & \equiv &
\langle y({\bf k},\eta_0),y(-{\bf k},\eta_0)|\alpha\beta,~\eta \rangle_S
\nonumber\\
& = & N_k~\exp \lbrack~
-\gamma_k~(y_0-\langle y \rangle)
(y^*_0-\langle y \rangle^*)+i(y_0^*~\langle p \rangle
+ y_0~\langle p \rangle^*)~\rbrack  \label{psialpha}\label{psialpha}\\
& = & N_k~\exp \lbrack~
-\gamma_k~|y_0-\langle y \rangle|^2 +i(~y_0^*~\langle p \rangle
+ y_0~\langle p \rangle^*~)~\rbrack~. \label{psialpha1}
\end{eqnarray}
We could equally well arrive at that result working in the real and imaginary
coordinate representation. Indeed, we can separate the variables $y$ and $p$
in real and imaginary part and consider the corresponding operators in the
Schr\"odinger representation. Defining
$S~a(\pm {\bf k},\eta_0)~S^{-1} \equiv a_1(\eta) \pm i a_2(\eta)$, the above
equations can be written as :
\begin{eqnarray}
a_1(\eta)~|\alpha\beta, \eta \rangle_S  & \equiv &
\lbrace g^*_k(\eta)~\langle y_1 \rangle + if^*_k(\eta)~\langle y_1 \rangle
\rbrace~|\alpha\beta,~\eta \rangle_S \nonumber \\
a_1(\eta)~\Psi_{\alpha\beta}(y_0,y_0^*,\eta) 
& = & g_k^*(\eta)~y_{01} + f_k^*(\eta)
\frac{1}{2}\frac{\partial}{\partial y_{01}}~
\Psi_{\alpha\beta}(y_0,y_0^*,\eta)~,\\
a_2(\eta)~|\alpha\beta,~\eta \rangle_S & \equiv &
\lbrace g^*_k(\eta)~\langle y_2 \rangle + if^*_k(\eta)~\langle y_2 \rangle
\rbrace~|\alpha\beta,~\eta \rangle_S \nonumber \\
a_2(\eta)~\Psi_{\alpha\beta}(y_0,y_0^*,\eta) 
& = & g_k^*(\eta)~y_{02} + f_k^*(\eta)
\frac{1}{2}\frac{\partial}{\partial y_{02}}~
\Psi_{\alpha\beta}(y_0,y_0^*,\eta)~.
\end{eqnarray}
Solving this differential system, we find that the wave function
can be written as: $\Psi_{\alpha\beta}(y_0,y_0^*,\eta) =
\phi(y_{01}, \eta) \phi(y_{02},\eta)$,
where :
\begin{equation}
\phi(y_{01}, \eta) = \sqrt{N_k}
\exp \lbrack~-\gamma_k~(y_{01}-\langle y_1 \rangle)^2
+i2~y_{01}~\langle p_1 \rangle~\rbrack~,
\end{equation}
in complete accordance with (\ref{psialpha},\ref{psialpha1}).

We are further interested in states which, at some initial time $\eta_0$,
contain some arbitrary number $N$ of particles with momenta
${\bf k}~\rm{and}~-{\bf k}$ and we denote these states by
$|N_{\bf k},N_{-{\bf k}},~\eta_0\rangle_S \equiv |N,~\eta_0\rangle_S$:
\begin{equation}
|N,~\eta_0\rangle_S =\frac{1}{N!}
a^{\dagger N} ({\bf k},\eta_0)
a^{\dagger N} (-{\bf k},\eta_0)~|0,~\eta_0 \rangle_S~.\label{N}
\end{equation}
We deal with modes initially deep inside the Hubble radius
($k\gg a'(\eta_0)/a(\eta_0)$) and use the WKB solution
for these modes.
The time-evolved state defined in (\ref{N}) becomes :
\begin{equation}
|N,~\eta \rangle_S =
\frac{1}{N!}
\left(S~a^\dagger ({\bf k},\eta_0)~S^{-1}\right)^N
\left(S~a^{\dagger}(-{\bf k},\eta_0)~S^{-1}\right)^N
|0,~\eta \rangle_S~.
\end{equation}
In the Schr\"odinger coordinate representation, the following differential
equation relating the
N-particles two-modes wave function to the vacuum two-modes wave
function is obtained:
\begin{equation}
\Psi_N (y_0,y^*_0,\eta) = \frac{1}{N!} f_k^{2N}
\left( \gamma_k^*~y_0 - \frac{\partial}{\partial y^*_0}\right)^N
\left( \gamma_k^*~y^*_0 - \frac{\partial}{\partial y_0}\right)^N
\Psi_0 (y_0,y^*_0,\eta)\label{deq}
\end{equation}
with
\begin{eqnarray}
\Psi_0 ( y_0,y_0^*,\eta) & = &
\left(\sqrt{\pi} |f_k|\right)^{-1}
\exp \left( - \gamma_k~ y_0~y_0^* \right)\nonumber\\
& = &
N_k \exp \left( - \gamma_k~y_0~y_0^* \right)~.
\end{eqnarray}
The action on $\Psi_0$ of the operator inside the right parenthesis can be
easily derived using the
relation $\gamma_k^* + \gamma_k = |f_k|^{-2}$. Indeed:
\begin{equation}
\Psi_N =
\frac{1}{N!} {\left(\frac{f_k}{f_k^*}\right)}^N~N_k \left(
\gamma_k^*~y_0 - \frac{\partial}{\partial y^*_0}\right)^N y^{*N}_0
\exp (- \gamma_k~y_0~y_0^*)~.
\end{equation}
We make further a binomial expansion of the first parenthesis :
\begin{equation}
\Psi_N =
\frac{1}{N!} \left(\frac{f_k}{f_k^*}\right)^N N_k
\sum_{p=0}^{N} {\cal C}_N^p
( \gamma_k^*~y_0 )^p
\left( -\frac{\partial}{\partial y_0^*} \right)^{N-p}
\left\{y_0^{*N} \exp \left( - \gamma_k~y_0~y_0^* \right) \right\}~.
\end{equation}
It is now convenient to use the following trick, namely to replace
$( \gamma_k^*~y_0 )^p$\ by the equivalent expression:
\begin{equation}
( \gamma_k^*~y_0 )^p =
\exp ( \gamma_k^*~y_0~y_0^* )
\left( -\frac{\partial}{\partial y_0^*} \right)^p
\left\{ \exp ( - \gamma_k^*~y_0~y_0^* ) \right\}
\end{equation}
As a result we get
\begin{eqnarray}
\Psi_N &=&
\frac{1}{N!}{\left(\frac{f_k}{f_k^*}\right)}^N N_k
\exp \left(\gamma_k^*~y_0~y_0^* \right)
\sum_{p=0}^{N} {\cal C}_N^p
\left( -\frac{\partial}{\partial y_0^*} \right)^p
\left\{ \exp \left(- \gamma_k^*~y_0~y_0^* \right) \right\} \times\nonumber\\
& &~~~~
\left( -\frac{\partial}{\partial y_0^*} \right)^{N-p}
\left\{ \exp (-\gamma_k~y_0~y_0^* ) \right\},
\end{eqnarray}
where we recognize the $N$th derivation of a product of functions.
The Leibnitz formula gives:
\begin{eqnarray}
\Psi_N &=&
\frac{1}{N!} \left( \frac{f_k}{f_k^*}\right)^N N_k
\exp ( \gamma_k^*~y_0~y_0^* )
\left( - \frac{\partial}{\partial y_0^*} \right)^N
\left\{ y_0^{*N} \exp ( -(\gamma_k^* + \gamma_k)~y_0~y_0^* ) \right\}
\nonumber\\
& = &
\frac{1}{N!} \left(-\frac{f_k}{f_k^*}\right)^N~\Psi_0
\left[ \exp \left(\frac{|y_0|^2}{|f_k|^2} \right)
\left( \frac{\partial}{\partial y_0^*} \right)^N
\left\{ y_0^{*N} \exp -\frac{{|y_0|}^2}{{|f_k|}^2} \right\} \right]~.
\end{eqnarray}
The expression inside the square brackets is just a Laguerre polynomial
as can be checked using Rodrigue's formula:
\begin{equation}
L_N (w) = \frac{1}{N!}e^w \left( \frac{\partial}{\partial w} \right)^N
\left\{ w^N e^{-w} \right\}~.
\end{equation}
Taking $w(y^*) = -\frac{y_0}{{|f_k(\eta)|}^2} y_0^*$, where $y_0$\ is
constant, it follows that the expression inside the square brackets is just
$L_N\left(\frac{y_0 y_0^*}{{|f_k(\eta)|}^2}\right)$.
The final result is therefore:
\begin{equation}
\Psi_N(y_0,y_0^*,\eta) = \left(-\frac{f_k}{f_k^*}\right)^N
L_N \left(\frac{|y_0|^2}{|f_k|^2}\right)
\Psi_0 (y_0,y_0^*,\eta)~.\label{N1}
\end{equation}
This calculation can also be done differently, in a way which takes
advantage of basic results pertaining to the familiar one-dimensional
harmonic oscillator. Indeed, let us separate the complex variables and
their corresponding operators in real and imaginary parts, as we did
already previously.
The definition of the state then becomes :
\begin{eqnarray}
|N,~\eta \rangle_S
& = &
\frac{1}{N!}
\left( S~a^\dagger ({\bf k},\eta_0)~S^{-1}~S~a^\dagger (-{\bf k},\eta_0)~
S^{-1}\right)^N |0,\eta \rangle_S \nonumber\\
& = &
\frac{1}{N!}
\left( a_1^{\dagger 2} + a_2^{\dagger 2} \right)^N |0,\eta\rangle_S\nonumber\\
& = &
\frac{1}{N!}\left(\sum_{M=0}^{N} {\cal C}_N^M
    a_1^{\dagger 2M} a_2^{\dagger 2N-2M} \right) |0,\eta\rangle_S~.
\end{eqnarray}
In this representation, the differential equation (\ref{deq}) becomes:
\begin{eqnarray}
\Psi_N(y_0,y_0^*,\eta) & = &
\frac{1}{N!} f_k^{2N}(\eta) \sum_{M=0}^{N} {\cal C}_N^M
\left( \gamma_k^*~y_{01} - \frac{1}{2}\frac{\partial}{\partial y_{01}}
\right)^{2M} \times \nonumber\\
& &~~~~\left( \gamma_k^*~y_{02} - \frac{1}{2}\frac{\partial}{\partial y_{02}}
\right)^{2N-2M} \Psi_0( y_0,y_0^*,\eta) \nonumber\\
& = &
\frac{N_k}{N!} f_k^{2N} \sum_{M=0}^{N} {\cal C}_N^M
\left( \gamma_k^*~y_{01} - \frac{1}{2}\frac{\partial}{\partial y_{01}} \right)
^{2M} \left\{e^{ - \gamma_k y_{01}^2}\right\}\times \nonumber\\
& &~~~~\left( \gamma_k^*~y_{02} - \frac{1}{2}\frac{\partial}{\partial y_{02}}
\right)^{2N-2M} \left\{e^{- \gamma_k~y_{02}^2} \right\}~.
\end{eqnarray}
Knowing the harmonic oscillator eigenfunctions, it is straightforward to find:
\begin{equation}
\Psi_N  = \frac{1}{N! 4^N}
\left(\frac{f_k}{f_k^*}\right)^N  \sum_{M=0}^{N} {\cal C}_N^M
H_{2M}\left( \frac{y_{01}}{|f_k|} \right)
H_{2N-2M}\left(\frac{y_{02}}{|f_k|} \right)
\Psi_0~.\label{N2}
\end{equation}
It is now possible to show that this expression is equal to (\ref{N1}).
Taking into account the relation between Hermite and Laguerre
polynomials, we write:
\begin{eqnarray}
\Psi_N  & = &
\frac{1}{N! 4^N}
{\left(\frac{f_k}{f_k^*}\right)}^N \sum_{M=0}^{N} {\cal C}_N^M
{(-1)}^{M} 2^{2M} M!
L_M^{(-\frac{1}{2})} \left(\frac{{y_0}_1^2}{|f_k|^2} \right) \times
\nonumber\\
& \: &
\qquad
{(-1)}^{N-M} 2^{2N-2M} (N-M)!
L_{N-M}^{(-\frac{1}{2})} \left(\frac{{y_0}_2^2}{|f_k|^2} \right)
\Psi_0 \nonumber\\
& = &
\frac{1}{N!}
{\left(-\frac{f_k}{f_k^*}\right)}^N \left(\sum_{M=0}^{N}
L_M^{(-\frac{1}{2})} \left(\frac{{y_0}_1^2}{|f_k|^2} \right)
L_{N-M}^{(-\frac{1}{2})} \left(\frac{{y_0}_2^2}{|f_k|^2} \right) \right)
\Psi_0 \nonumber\\
& = &
\frac{1}{N!} \left(-\frac{f_k}{f_k^*}\right)^N
L_N \left(\frac{y_{01}^2 + y_{02}^2 }{|f_k|^2} \right) \Psi_0~,
\end{eqnarray}
where 8.974.4, p.1038 of~\cite{GR} has been used in the last step of the
above calculation.

The two expressions (\ref{N1},\ref{N2}) are two equivalent ways to write
down the wave function in the coordinate Schr\"odinger representation. They
represent, together with (\ref{psialpha}), the basic result of this section.
We note that the wave function (\ref{N1}) is a function of $|y(\bf{k})|$
and therefore, like the vacuum wave function, it is invariant under
reflection $\bf{k}\to -\bf{k}$.
As a result, it is clear that
all correlation with an odd number of points will vanish, the distribution
has no skewness.
On the other hand, it is of course no more a Gaussian, a fact which
will be important regarding the positivity of the Wigner function.
The transition to semi-classicality can now be seen in this representation.
The condition for a semiclassical behaviour is that
the phase of the wave function should be large for a typical value of
$|y|^2\sim \langle |y|^2\rangle =(2N+1)|f_k|^2$ . This results in the 
inequality $(2N+1)|F_k(\eta)|\gg \hbar$ (here we restored $\hbar$ for 
clarity). Note that, as one could expect, the latter condition is more
easily achieved for large $N$. The
probability distribution for $y_0$  at time $\eta$ is given by:
\begin{equation}\label{distr}
\rho_N (|y_0|, \eta)= \left| \Psi_N (y_0,y_0^*,\eta) \right|^2
=
\frac{1}{\pi |f_k|^2}~
L_N^2 \left( \frac{|y_0|^2}{|f_k|^2}\right)
\exp \left( - \frac{|y_0|^2}{|f_k|^2} \right)~. \label{prob}
\end{equation}
This probability distribution conserves probability along the classical 
trajectory $y\propto f_k(\eta)$ when the condition $(2N+1)|F(k)|\gg 1$ is 
satisfied.
As we shall show below in section 4, the requirement that the latter condition
should be better and better satisfied with the expansion of the Universe
picks up a small subset of all possible classical trajectories $f_k(\eta)$ in 
phase space (namely, the quasi-isotropic mode of perturbations).

\section{Heisenberg representation}

Let us perform now an independent analysis of the transition to 
semiclassical behaviour in the Heisenberg representation.
The crucial point here is that, as shown in~\cite{cqg96}, in the limit
of infinite squeezing the mode functions $f_k(\eta)$ can be made real by a 
time-independent phase rotation (of course, in addition, $|f_k|\to \infty$).
Then the momentum mode functions $g_k(\eta)$ become purely imaginary. 
As a result, the field operators $y({\bf k},\eta)$ and $p({\bf k},\eta)$
take the following form:
\begin{eqnarray}
y({\bf k},\eta) &=& f_k~\left( a({\bf k},\eta_0)+a^\dagger (-{\bf k},\eta_0)
\right)~,\\
p({\bf k},\eta) &=& -ig_k~\left( a({\bf k},\eta_0)+a^\dagger (-{\bf k},\eta_0)
\right)=-i~{g_k\over f_k}~y({\bf k},\eta)~. \label{trans}
\end{eqnarray}
So, in the limit $|r_k|\to \infty$, the non-commutativity of 
$y({\bf k},\eta)$ and $p({\bf k}, \eta)$ (and of all other operators, too) 
may be neglected. Due to this reason, we can introduce equivalent classical 
stochastic quantities $y_k(\eta)=f_k~e({\bf k})$ where $e({\bf k})$ are
c-number complex stochastic 
time-independent variables obeying the same statistics
as $y({\bf k},\eta_0)$ at the time $\eta_0$. They have zero average and unit
dispersion: $\langle e({\bf k})\rangle =0,~~\langle e({\bf k})e^*({\bf k}'
\rangle =\delta^{(3)}({\bf k}-
{\bf k}')$, though they are generally not Gaussian. The definition of the
equivalence is the following: we say that a complex classical stochastic 
field 
$y(\eta)$ with a corresponding momentum $p(\eta)$ and a probability
distribution in phase space $\rho (|y|,|p|)$ are equivalent
to quantum operators $\hat y(\eta )$ and $\hat p(\eta)$ if they give
the same average values for an arbitrary function $G(y,p)$:
\begin{equation}
_H \langle N| G(\hat y, \hat p)G^\dagger(\hat y, \hat p)|N \rangle_H =
\int \int dy_1 dy_2 dp_1 dp_2\rho(|y|,|p|) \left| G(y,p) \right|^2~
\label{equiv}
\end{equation}
(here we omit ${\bf k}$ for brevity but we restore hats over operators).

Of course, such a condition cannot be fulfilled generally, since quantum 
mechanics is not equivalent to classical field theory. However, in our case
of infinite squeezing, due to relation (\ref{trans}) between $\hat y$ and 
$\hat p$ (while the same relation between classical stochastic quantities
$y$ and $p$ holds), eq. (\ref{equiv}) reduces to a much simpler form: 
\begin{equation}
_H \langle N| \tilde G(y)\tilde G^\dagger(y)|N \rangle_H =
\int \int dy_1 dy_2 \rho(|y|) \left| \tilde G(y) \right|^2 \label{heis}
\end{equation}
where $\tilde G(y)=G(y,~p(y))$ (since all operators are mutually commuting,
we do not write any more hats over them). For the $N$-particles initial state, 
the probability density function $\rho(|y|)$ is given by
\begin{equation}\label{distr1}
\rho(|y|)= \frac{1}{\pi |f_k|^2} L_N^2 \left(\frac{|y|^2}{|f_k|^2}\right)
\exp \left( - \frac{|y|^2}{|f_k|^2} \right)~.
\end{equation}
So, in the limit of infinite squeezing and {\em after} use is made of relation
(\ref{trans}) the equivalence condition takes a form which is identical to 
the {\em definition} of an average value of an operator in quantum mechanics
with the probability distribution $\rho$ equal to the modulus square of a
wave function satisfying the Schr\"odinger equation. Thus, (\ref{heis})
can be immediately
proved by referring to the results of the previous section. However, if we
want to produce a completely independent derivation in the Heisenberg 
representation, we should not use the fact that $\rho$ is determined by
some solution of the Schr\"odinger equation but instead we have to make an 
actual calculation of both sides of eq. (\ref{heis}) and prove their equality.
 
Further, we omit tilda over $G$. We compute first the l.h.s. of (\ref{heis}):
\begin{equation}
_H \langle N |G(y)G^\dagger (y) |N\rangle_H =
\sum_{m, n=0}^\infty q_m q_n^*~_H \langle N|
y^m({\bf k},\eta) y^{\dagger n}({\bf k},\eta) |N\rangle_H~.
\end{equation}
We use now the expression (\ref{trans}) in the high squeezing limit when
$f_k(\eta)$ can be made real by a time-independent phase rotation, and write:
\begin{eqnarray}
_H \langle N| G(y)G^\dagger(y) |N\rangle_H =
\sum_{m, n=0}^\infty q_m q_n^*\frac{f_k^{m+n}}{(N!)^2}
_H \langle 0|a^N (-{\bf k},\eta_0)~a^N ({\bf k},\eta_0)~\times\nonumber\\
\left( a ({\bf k},\eta_0)+a^{\dagger} (-{\bf k},\eta_0) \right)^m
\left( a^{\dagger}({\bf k},\eta_0) + a (-{\bf k},\eta_0) \right)^n
a^{\dagger N}({\bf k},\eta_0) a^{\dagger N}(-{\bf k},\eta_0) | 0 \rangle_H~.
\end{eqnarray}
We find after a somehow tedious but straightforward calculation:
\begin{equation}\label{left}
{\:}_H \langle N| G(y)~G^\dagger(y) |N\rangle_H =
\sum_{m=0}^\infty {|q_m|}^2 f_k^{2m}(\eta)
\sum_{i=sup(0, m-N)}^{m}
\frac{{m!}^2 (N+i)!}{{i!}^2 {(m-i)!}^2 (N-m+i)!}~.
\end{equation}
As for the r.h.s. of (\ref{heis}), we get:
\begin{eqnarray}
& \: &
\int \!\!\! \int dy_{1}dy_{2}\:
\rho(|y|)| G(y) |^2
\nonumber\\
& = &
\!\!\!\!\!\!\!\! \sum_{m, n=0}^\infty q_m q_n^*
\int \!\!\! \int dy_1 dy_2
\frac{1}{\pi |f_k|^2} \:
{ L_N\left(
\frac{{|y|}^2}{|f_k|^2}\right) }^2y^m y^{*n}
\exp \left( - \frac{{|y|}^2}{|f_k|^2} \right)~.
\end{eqnarray}
Replacing $L_N$\ by its explicit polynomial expression, we find:
\begin{eqnarray}\label{right}
& \: &
\int \int dy_1dy_2\:
\rho \left( |y|^2 \right) \left| G\left(y \right) \right|^2 \nonumber\\
& = &
\sum_{m=0}^\infty {|q_m|}^2 f_k^{2m}
\sum_{i,j = 0}^{N}
\frac{ {(-1)}^{i+j} {N!}^2}
     { {i!}^2 {j!}^2 (N-i)! (N-j)!}
\int_0^\infty x^{m+i+j} e^{-x} dx
\nonumber\\
& = &
\sum_{m=0}^\infty {|q_m|}^2 f_k^{2m}
\sum_{i, j = 0}^{N}
\frac{ {(-1)}^{i+j} {N!}^2 (m+i+j)!}
     { {i!}^2 {j!}^2 (N-i)! (N-j)!}~.
\end{eqnarray}
One finally checks, as we did numerically, that the sums in (\ref{left})
and (\ref{right}) are indeed equal:
\begin{equation}
\sum_{i=sup(0, m-N)}^{m}
\frac{{m!}^2 (N+i)!}{{i!}^2 {(m-i)!}^2 (N-m+i)!} =
\sum_{i, j = 0}^{N}
\frac{ {(-1)}^{i+j} {N!}^2 (m+i+j)!}
     { {i!}^2 {j!}^2 (N-i)! (N-j)!}~.
\end{equation}
This ends the proof of equality (\ref{heis}).
We note that the
calculation done above yields in particular all the correlation functions.
Indeed, if we set $G(y) = y^m$, we get:
\begin{equation}
{\:}_H \langle N| \left|y\right|^{2m} | N\rangle_H =
|f_k|^{2m} \sum_{i=sup(0, m-N)}^{m}
\frac{{m!}^2 (N+i)!}{{i!}^2 {(m-i)!}^2 (N-m+i)!}~.
\end{equation}
We see also that all the following correlation functions do vanish:
\begin{equation}
{\:}_H \langle N| y^m y^{\dagger n} |N\rangle_H = 0 \qquad \mbox{if}
\qquad m \neq n~.
\end{equation}
Having in mind the comparison of the probability density obtained here with
a Gaussian one, the two-point and four-point correlation functions are
of particular importance:
\begin{eqnarray}
{\:}_H \langle N| \left| y \right|^2
| N\rangle_H
& = &
(2N + 1) |f_k|^2 \label{2p}~, \nonumber \\
{\:}_H \langle N | \left| y \right|^4 | N\rangle_H
& = &
(6N^2 + 6N + 2) |f_k|^4~.\label{4p}
\end{eqnarray}
The above results can be directly used in order to calculate the kurtosis
$Q$ (the irreducible 4-order moment) of the probability distribution:
\begin{equation}
Q\equiv \frac{\langle N,~\eta||y({\bf k},\eta_0)|^4|N,~\eta \rangle}
{\langle N,~\eta||y({\bf k},\eta_0)|^2|N,~\eta \rangle^2}-2 =
-\frac{1}{2} \left(1- {1\over (2N+1)^2}\right)~.
\end{equation}
It vanishes, as should be the case, when $N=0$ and tends to $-1/2$ for
$N\to \infty$. We will come back to this point in Section 6 and in the
conclusion.
For comparison, we present the corresponding results for a one-mode 
$N$-particles state for which
\begin{equation} \label{A1}
\varphi =\varphi(t)a+\varphi^{\ast}(t)a^{\dagger},\qquad [a,a^\dagger]=1~.
\end{equation}
An example of such a state is the ${\bf k}=0$\ mode.
Then
\begin{eqnarray} \label{A2}
\langle \varphi^2 \rangle &=& (2N+1) |\varphi|^2~, \nonumber \\
\langle \varphi^4 \rangle &=& (6N^2+6N+3) |\varphi|^4~, \nonumber \\
Q &\equiv& \frac{\langle \varphi^4\rangle}{\langle \varphi^2\rangle^2}-3
=-\frac{3}{2}\left(1-\frac{1}{(2N+1)^2}\right)~\label{onemode}.
\end{eqnarray}
These expressions will be also useful for the calculation of the curtosis 
of $y$ in the coordinate representation, see section 7. 

\section{Conditions for semiclassical behaviour of perturbations}

Let us now analyze in more detail the conditions imposed on the 
mode functions $f_k(\eta)$ of the quantum field $y$ or on the equivalent 
stochastic functions $y_k(\eta)=f_k(\eta)e({\bf k})$ by the requirement of
high squeezing $|r_k|\rightarrow\infty$\ which is crucial for the
quasi-classical equivalence as we have shown in sections 2 and 3.
In the long-wavelenght limit $k \ll aH$, the leading term in the mode 
functions $f_k(\eta)$ and
$g_k(\eta)$\ have the form \cite{staro79,cqg96}:
\begin{equation} \label{A3}
f_k=C_1({\bf k})a+C_2({\bf k})a\int_{\infty}^{\eta}\frac{d\eta^{\prime}}
{a^2(\eta^{\prime})}, \qquad
g_k={\cal O}(iC_1({\bf k})\frac{k^2}{H})+\frac{iC_2({\bf k})}{a}
\end{equation}
where $C_1({\bf k})$\ is made real and positive by a phase rotation. 
We assume here that the integral converges at the lower
limit that takes place for all physically interesting equations of state
of matter at late times, in particular, if $a(t)\propto t^n$ with $n>1/3$).
$C_1({\bf k})$\ and
$C_2({\bf k})$\ should satisfy the normalisation condition
\begin{equation} \label{A4}
C_1 \Im C_2 = -\frac{1}{2}~.
\end{equation}
The first term in (\ref{A3}) is the quasi-isotropic mode, the second is the 
decaying mode. 
In the simplest inflationary models with one slow-rolling scalar 
field, a more concrete expression for $C_1({\bf k})$\ and $C_2({\bf k})$ 
is obtained~\cite{staro79,cqg96}:
\begin{equation} \label{A5}
C_1= \frac{H_k}{\sqrt{2} k^{3/2}}, \qquad C_2= -\frac{ik^{3/2}}{\sqrt{2}H_k}
\end{equation}
where $H_k$\ is the value of $H(\eta)$\ at the moment of the first Hubble 
radius crossing during the inflationary stage. However, generally we need 
not restrict ourselves to this case only. The first term in the expression 
for $g_k$ in eq.(\ref{A3}) has a rather complicated integro-differential 
structure that can 
be defined by solving (\ref{weq}) using a perturbative expansion in powers of 
$k^2$. The given estimate is valid for a wide class of smooth functions 
$a(\eta)$ including arbitrary power-law and quasi-de Sitter ones.

Now let us insert (\ref{A3}) into the definition of $F(k)$, eq.(\ref{defF}). 
It is argued in \cite{cqg96} that it is the condition $|F(k)|\gg1$\ that is 
required for the quasi-classical behaviour of the vacuum initial state. 
In the case of $N$-particles state, the necessary condition is weaker, 
namely $(2N+1)|F(k)|\gg 1$ as shown above. Assuming a power-law or a
quasi-de Sitter behaviour of the scale factor $a(\eta)$
and omitting numerical coefficients of the order
of unity, we obtain the following condition:
\begin{equation} \label{A6}
\left| C_1^2 \frac{k^2 a}{H}+\frac{|C_2 |^2}{a^3 H}+C_1 \Re C_2 \right| \gg 
\frac{1}{2N+1}~.
\end{equation}
If the solution is dominated by the quasi-isotropic mode, and the term with 
$C_2$ may be neglected, the condition (\ref{A6}) takes the form
\begin{equation} \label{A7}
C_1^2\gg\frac{H}{k^2a}\cdot\frac{1}{2N+1}~.
\end{equation}
In terms of metric tensor perturbations (gravitational waves) which are 
related to $\varphi=y/a$ by the relation
\begin{equation} \label{A8}
h_{ij}\equiv -\frac{\delta g_{ij}}{a^2}=\sqrt{32 \pi G}\: \varphi\: e_{ij}
\end{equation}
where $e_{ij}({\bf k})$\ is the polarisation tensor normalised by the 
condition $e_{ij}e^{ij}=1, \quad i,j=1,2,3$, we get:
\begin{equation} \label{A9} 
k^3 h_g^2(k)\equiv \langle k^3 h_{ij}({\bf k}) h^{ij \ast} ({\bf k}) \rangle 
\gg \frac{H\, l_P^2}{\lambda} \cdot \frac{1}{2N+1}
\end{equation}
where $l_P=\sqrt{G}$\ is the Planck length and $\lambda=2 \pi a k^{-1}$
is the wavelength of the perturbation.

The condition (\ref{A7}) is increasingly better satisfied with the 
expansion of the Universe. Thus, the quasi-isotropic mode becomes more 
and more classical at later times. For the simplest inflationary initial 
conditions (\ref{A5}), the inequalities (\ref{A7}) and (\ref{A9}) are 
satisfied if $\lambda H (2N+1) \gg 1$, in particular, everywhere in the 
long-wavelenght regime $\lambda H \gg 1$\ even for $N=0$.

Eqs. (\ref{A7}) and (\ref{A9}) were derived under the assumption 
$\lambda H \gg 1$. In the opposite case $\lambda H \ll 1$,
i.e. after the last Hubble crossing, eq. (54) of paper \cite{cqg96} must 
be used instead of (\ref{A3}). This leads to the classicality condition:
\begin{equation} \label{A10}
k^3 h_g^2(k) \gg \frac{l_P^2}{{\lambda}^2} \cdot \frac{1}{2N+1}
\end{equation}
which was derived in \cite{cqg96} for the case $N=0$.
On the other hand, let us consider the case where the quasi-isotropic mode is 
negligible with respect to the decaying mode while the latter is 
imaginary ($|\Re C_2 | \ll | \Im C_2 |$\ in the representation where 
$\Im C_1 = 0$).
This requires, in particular, $C_1^2$\ to be much less than the value 
given by (\ref{A5})\ at the moment of any Hubble radius crossing. Note that 
$\Im C_2$\ is uniquely defined by $C_1$ through the normalisation condition 
(\ref{A4}) while $\Re C_2$\ is not restricted at all. Then the classicality 
condition takes the form:
\begin{equation} \label{A11}
| C_2 |^2 \gg \frac{a^3 H}{2N+1}, \qquad k^3 h_g^2(k) \gg \frac{l_P^2}
{{\lambda}^3 H} \cdot \frac{1}{2N+1}~.
\end{equation}
These inequalities become weaker with the expansion of the Universe for all 
interesting cases (in particular, for $a(t) \propto t^n$\ with $n > 
\frac{1}{3}$) and may even cease to be valid. Thus, the imaginary decaying 
mode becomes less classical and more quantum at later times.\\ 
Finally, a sufficiently large real and positive decaying mode ($\Re C_2 \gg
| \Im C_2 | = \frac{1}{2 C_1}$) is always classical, since in this case the 
inequality (\ref{A6}) is already satisfied due to the third term in the 
left-hand side. In contrast, a negative decaying mode ($\Re C_2 < 0$, $| \Re
C_2 | \gg | \Im C_2|$) has a rather peculiar behaviour: being classical at 
early and late times, it ceases to be classical at the moment when the first 
and the third term in the left-hand side of (\ref{A6}) almost cancel each 
other.

All that was said above refered either to the case of a massless minimally 
coupled scalar field $\varphi$, or to tensor metric perturbations 
(gravitational waves). In the case of scalar (or adiabatic) metric 
perturbations, one should first either specify a gauge, or choose some 
gauge-independent quantity describing the perturbations. Let us consider the 
gauge-independent quantity $\Phi$\ which coincides with the Newtonian 
potential in the longitudinal gauge ($g_{00} = 1+2 \Phi$). However, one should 
keep in mind that semiclassical behaviour may be more pronouced in terms of 
some quantities and less pronouced, or even absent, for some other quantities.

Let us further restrict ourselves to the case where pressure perturbations are 
diagonal. This includes all cases where matter is a mixture of an arbirary 
number of scalar fields with arbitrary potentials (but minimally coupled to 
the gravitational field) and an arbitrary number of ideal fluids. Then, as is 
well known, in the long-wave regime $\lambda \gg {\lambda}_J \sim 
H^{-1}c_S$\ where ${\lambda}_J$\ is the Jeans lenght and $c_S$\ is the 
sound velocity, a Fourier mode of the gravitational potential $\Phi_k$\ may 
be written as:
\begin{equation} \label{A12}
\Phi_k = \tilde{C}_1({\bf k}) \left( 1-\frac {H}{a} \int_{{\eta}_1}^{\eta}
a^2 d \eta \right) - \frac{4 \pi G}{k^2} \tilde{C}_2 ({\bf k}) \frac{H}{a}
\end{equation}
where $\tilde{C}_1({\bf k})$\ can be made real and positive by a phase 
rotation (see, e.g., \cite{david92} for a simple derivation of a more 
general statement valid in the synchronous gauge). Once more, the first 
term in the right-hand side of (\ref{A12}) is the quasi-isotropic mode (which 
may be really called the growing mode now since the corresponding density 
perturbation increases with the expansion of the Universe), the second term 
is the decaying mode. $\eta_1$\ should be as small as possible, but not less 
than the moment of the first Hubble radius crossing at the inflationary 
stage. 

If the operator $\Phi({\bf k},\eta)$\ is represented as:
\begin{equation} \label{A13}
Phi({\bf k},\eta) =
{\Phi}_k (\eta)~a({\bf k},{\eta}_0) +
{\Phi}_k^{\ast}(\eta)~a^{\dagger}(-{\bf k},{\eta}_0)~,
\end{equation}
then the normalisation condition for the coefficients 
$\tilde{C}_1 ({\bf k})$\ and
$\tilde{C}_2 ({\bf k})$\ of the mode functions ${\Phi}_k (\eta)$\ in the 
long-wave regime has the same form as (\ref{A4}):
\begin{equation} \label{A14}
{\tilde{C}}_1 \Im {\tilde{C}}_2 = -\frac{1}{2}~.
\end{equation}
For the case of matter in a Universe consisting of one minimally coupled 
scalar field $\varphi$, this statement follows from the fact that in this 
case, canonical 
variables analogous to $y$\ and $p$\ are given by~\cite{muk88}:
\begin{equation} \label{A15}
\xi \equiv a \zeta = a \delta {\varphi}_L + \frac{{\varphi}^{\prime}}{H} 
\Phi~,~~~{\pi}_{\xi} = {\xi}^{\prime} - \frac{a^{\prime}}{a} \xi
\end{equation}
where $\delta {\varphi}_L$\ is the scalar field perturbation in the 
longitudinal gauge. Using two exact relations between $\zeta_k$\ and 
$\Phi_k$:
\begin{eqnarray} \label{A16}
\Phi_k &=& - \frac{4 \pi G a^2}{k^2} \: \frac{\dot{\varphi}^2}{H} \:
\left( \frac{H}{\dot{\varphi}} \zeta_k \right)^\cdot; \nonumber \\
\zeta_k &=& \frac{1}{4 \pi G} \: \frac{H}{a \dot{\varphi}} \:
\left( \frac{a}{H} \Phi_k \right)^{\cdot} 
\end{eqnarray}
and the exact background equation $\dot{H} = - 4 \pi G \: {\dot{\varphi}}^2$,
it is straightforward to obtain the leading terms in ${\zeta}_k$\ and 
${\xi}_k$\ in the long-wavelenght region from eq. (\ref{A12}):
\begin{eqnarray} \label{A17}
{\zeta}_k &=& {\tilde{C}}_1 \: \frac{\dot{\varphi}}{H} +
{\tilde{C}}_2 \: \frac{\dot{\varphi}}{H} \: \int_\infty^t 
\frac{H^2}{{\dot{\varphi}}^2 a^3} dt\: ;  \nonumber \\
\xi_k &=& {\tilde{C}}_1 \: \frac{{\varphi}^{\prime}}{H} +
{\tilde{C}}_2 \: \frac{{\varphi}^{\prime}}{H} \: \int_\infty^\eta 
\frac{H^2}{\varphi^{\prime 2}} d \eta~.
\end{eqnarray}
Since, on the other hand, $\xi_k$\ should satisfy the normalisation condition 
$\xi_k^\prime \, \xi_k^\ast - \xi_k^{\prime \ast} \, \xi_k = -i$, we arrive to 
(\ref{A14}). Note also the exact equal-time commutation relation following 
from the canonical commutation relations between $\xi$\ and $\pi_\xi$ and 
eq. (\ref{A16}):
\begin{equation} \label{A18}
\left[ \Phi ({\bf k},\eta), \dot{\Phi}^\dagger ({\bf k},\eta) \right] =
-i \frac{4 \pi G}{k^2} \: \frac{\dot{H}}{a}~.
\end{equation}

For the simplest inflationary models driven by one slow-rolling minimally 
coupled scalar field $\varphi$:
\begin{equation} \label{A19}
\tilde{C}_1 ({\bf k}) = 
\frac{H_k^2}{\sqrt{2} k^{3/2} |\dot{\varphi}_k| }
\quad , \quad
\qquad 
\tilde{C}_2 ({\bf k}) = -\frac{i k^{3/2} |\dot{\varphi}_k|}{\sqrt{2} H_k^2}
\end{equation}
where $H_k$\ and $\varphi_k$\ are the values of $H(t)$\ and $\varphi(t)$\ 
at the moment of the first Hubble radius crossing during the inflationnary 
stage. Eq. (\ref{A19}) leads to the standard expressions for density 
perturbations generated during inflation.

Let us now turn to the case of matter consisting of one hydrodynamical 
component with an ideal fluid equation of state $P=P(\varepsilon)$, 
where $P$\ is the pressure, $\varepsilon$\ is the energy density. The 
canonical variables analogous to $y$\ and $p$\ in this case were first 
found by Lukash \cite{luk80}:
\begin{eqnarray} \label{A20}
\bar{q} \equiv \bar{a}q &=& -\frac{3 \bar{a}}{\sqrt{8 \pi G}} 
\frac{H^2}{a \dot{H}} \: \left( \frac{a}{H} \Phi \right)^\cdot,
\quad \bar{a}= \frac{a}{c_S} \: \sqrt{\frac{1+\frac{P}{\varepsilon}}{3}},
\quad c_S^2 = \frac{dP}{d\varepsilon}, \nonumber \\
\pi_{\bar{q}} &=& \bar{q}^\prime \: - \: \frac{\bar{a}^\prime}{\bar{a}} 
\bar{q}
\end{eqnarray}
(it is assumed here that $c_S \neq 0$).
If the sound velocity $c_S$\ is constant, then $\bar{a} \propto a$.
The corresponding wave equation for phonon mode functions has the form:
\begin{equation} \label{A21}
\bar{q}_k^{\prime \prime} +
\left( k^2 c_S^2 - \frac{\bar{a}^{\prime \prime}}{\bar{a}} \right) \: 
\bar{q}_k = 0~.
\end{equation}
Note also the second relation between $q_k$\ and $\Phi_k$:
\begin{equation} \label{A22}
\Phi_k = - \frac{3 \sqrt{2 \pi G}}{k^2} \: H \, {\bar a}^2 \, \dot{q}_k~,
\end{equation}
and two useful identities for the background metric in this case:
\begin{equation} \label{A23}
1+\frac{P}{\varepsilon} = - \frac{2 \dot{H}}{3 H^2} 
\quad , \quad
1+c_S^2 = -\frac{\ddot{H}}{3 H \dot{H}}~.
\end{equation}

The corresponding leading long-wavelength behaviour of $q_k$\ is:
\begin{equation} \label{A24}
q_k = \frac{3}{\sqrt{8 \pi G}} \: \left( \tilde C_1 +
\frac{8 \pi G}{3} \: \tilde C_2 \int_\infty^t \frac{dt}{a^3} \cdot 
\frac{c_S^2}{1+\frac{P}{\varepsilon}} \right)~. 
\end{equation}
Using (\ref{A20})\ and (\ref{A22}), it is easy to verify that the commutation 
relation (\ref{A18})\ holds in this case too, which leads to (\ref{A14})\ 
once more. Though we derived the commutation relation (\ref{A18})\ for two 
particular cases (one scalar field or one ideal fluid), we believe that it 
has a much wider range of validity and we leave the proof of this assertion 
for another publication. 

The condition for semiclassical behaviour of the gravitational potential is:
\begin{equation} \label{A25}
\left| \langle \Phi({\bf k},\eta)~\dot{\Phi}^\dagger({\bf k},\eta)+ c.c. 
\rangle \right| \gg
\left| [ \Phi({\bf k},\eta)~,~\dot{\Phi}^\dagger({\bf k},\eta) ] \right|~.
\end{equation}
Substituting here the long-wave asymptote (\ref{A12}) and assuming 
hydrodynamical matter with a constant sound velocity (that leads to the 
power-law behaviour $a(t)\propto a^n$ with ${1\over 3}\le n \le {2\over 3}$),
we get the following condition of classicality similar to (\ref{A6}):
\begin{equation} \label{A26}
\left| \tilde{C_1}^2\cdot {\cal O}(1)\, \frac{k^2}{4\pi G}\, 
\frac{k^2c_S^2}
{aH^3}+|\tilde{C_2}|^2 \,(n+1)\, \frac{4 \pi G}{k^2} \, \frac{H}{a} \, 
- \tilde{C_1} \Re \tilde{C_2}\right|
\gg \: {1\over 2N+1}~.
\end{equation}
If the growing mode is dominating, then
\begin{equation} \label{A27}
\tilde{C_1}^2 \gg \frac{4 \pi G}{k^2} \, \frac{aH^3}{k^2c_S^2} \, 
\frac{1}{2N+1}~.
\end{equation}
This condition is increasingly better satisfied with the expansion of the 
Universe. In terms of the gravitational potential, this leads to
\begin{equation} \label{A28}
k^3 \Phi^2 (k) \equiv \langle k^3 | \Phi ({\bf k}) |^2 \rangle \gg \frac{l_P^2
\lambda H^3}{c_S^2}\, \frac{1}{2N+1}~. 
\end{equation}
On the other hand, for a purely decaying imaginary mode ($|\Re\tilde{C}_2|\ll 
|\Im \tilde{C}_2|$), we get
\begin{equation} \label{A29}
|C_2|^2 \gg \frac{k^2}{4 \pi G} \, \frac{a}{H}\frac{1}{2N+1}
\quad , \quad
k^3 \Phi^2 (k) \gg \frac{l_P^2H}{\lambda}\frac{1}{2N+1}~.
\end{equation}
This condition weakens with the expansion of the Universe. Note that the 
conditions for the gravitational potential (\ref{A28},\ref{A29}) are 
more restrictive than the corresponding conditions for gravitational waves
(\ref{A9},\ref{A11}) since the right-hand sides of the former conditions 
are in $(\lambda H)^2 \gg 1$ times larger (however, the growing mode is always 
semiclassical in the long-wavelength regime for the most interesting case
of perturbations produced during inflation with the initial conditions
(\ref{A19})). As pointed above, this peculiar feature may be a 
result of $\Phi$ not being the best quantity to study quantum-to-classical
transition.

In conclusion, we see that the condition for a perturbation to become more 
and more classical at later times (i.e. with the expansion of the Universe) 
chooses a subclass of all solutions for perturbations, namely, the 
quasi-isotropic modes. We will return to this point at the end of the paper. 

\section{The Wigner function}

The Wigner function was proposed as a candidate for the probability density
function of a quantum mechanical system in phase space. According to the
rules of quantum mechanics such a probability density can not exist, so
one can only hope to come as close as possible to it by requiring from
good candidates, like the Wigner function, to fulfill some basic 
properties of the probability density, but not all of them. In particular,
the Wigner function is real, but not definite positive. Let us
calculate the Wigner function for the system under consideration :
\begin{eqnarray}
W_{NN} &\equiv&
W\left( y_0,y_0^*,p_0,p_0^*\right)
\nonumber\\
&=&
\frac{1}{(2\pi)^2}~\int~\int dx_1~dx_2~e^{-i({\bar p}_1 x_1 + {\bar p}_2 x_2)}
\Psi_N^* \left(y-\frac{x}{2},\eta\right)
\Psi_N \left(y+\frac{x}{2},\eta\right)~.
\end{eqnarray}
We will use the simplified notations $ y({\bf k},\eta_0) \equiv y =
y_1 + i y_2 $ and the same convention for $ p({\bf k},\eta_0).$
The quantity ${\bar p}_1~,{\rm resp.}~{\bar p}_2$, is canonically conjugate
to $y_1~,{\rm resp.}~y_2$, and so ${\bar p}_1=2~p_1,~{\bar p}_2=2~p_2$.

We consider first the Wigner function $W_{\alpha\beta}$ of the coherent 
initial state
$\Psi_{\alpha\beta}$. Actually, it is clear from (\ref{psialpha}) that it
will be similar to the Wigner function $W_{00}$ of the vacuum
$|0,~\eta \rangle$. The following result is obtained
\begin{eqnarray}
W_{\alpha\beta} & = & N_k^2~e^{-\frac{|y_0-\langle y \rangle|^2}{|f_k|^2}}~
\frac{|f_k|^2}{\pi\hbar^2}~e^{-\frac{4|f_k|^2}{\hbar^2}
|\frac{F(k)}{|f_k|^2}(y-\langle y \rangle)-(p-\langle p \rangle)|^2 }
\nonumber \\
& \to & | \Psi_{\alpha \beta}|^2~\delta~\left( p_1-\frac{F(k)}{|f_k|^2}~y_1
\right)~~\delta~\left( p_2-\frac{F(k)}{|f_k|^2}~y_2 \right)~. \label{coh}
\end{eqnarray}
We recognize the vacuum Wigner function $W_{00}~{\rm with}~p\to p-\langle
p \rangle~{\rm and}~y\to y-\langle y \rangle$. 

We now turn our attention to
the Wigner function $W_{NN}$ of $N$-particles states. Our method will be to
compute the generating functional of $W_{NN}$. For this purpose we first 
write down the generating functional of $W_{mn}$:
\begin{eqnarray}
& & \sum_{m, n = 0}^{\infty} W_{mn} z^m v^n =
\frac{1}{(2\pi)^2}~\int~\int
dx_1~dx_2~e^{-i({\bar p}_1 x_1 + {\bar p}_2 x_2)}
\times
\nonumber\\
& &
\sum_{m, n = 0}^{\infty}
L_m \left( \frac{ {|y-x/2|}^2 }{ |f_k|^2 } \right)
L_n \left( \frac{ {|y+x/2|}^2 }{ |f_k|^2 } \right)
z^m z^n
\Psi_0^* \left(y-\frac{x}{2},\eta\right)
\Psi_0 \left(y+\frac{x}{2},\eta\right)~.\label{gencoh}
\end{eqnarray}
The generating functional of the Laguerre polynomials is given by:
\begin{equation}\label{genlag}
\sum_{m=0}^{\infty} L_m(x) z^m =
\frac{1}{1-z} \exp(-\frac{xz}{1-z}) \qquad \qquad
\left( |z| < 1 \right)~.
\end{equation}
We insert the expression for the vacuum wave function into (\ref{gencoh}),
replace each sum of Laguerre polynomials by the generating function
(\ref{genlag}) and rearrange terms in the
exponentials. The integration reduces to Fourier transforms
of gaussians. In order to give a compact result, we will set:
\begin{eqnarray}
Y =  \frac{y}{|f_k|}~, ~~~
P =  \left( 2 \frac{F_k}{|f_k|} y - |f_k| {\bar p} \right)~.
\end{eqnarray}

The result of the integration can then be written in the following way:
\begin{eqnarray}\label{finform}
\sum_{m, n = 0}^{\infty} W_{mn} z^m v^n &=&
\frac{1}{\pi^2} \frac{1}{1-zv}
\exp\left(-\frac{(|Y|^2 + |P|^2)(1+zv)}{1-zv} \right)
\times
\nonumber\\
& &
\exp\left(\frac{(-|Y|^2 + |P|^2 + 2i Y_1 P_1 + 2i Y_2 P_2)v}{1-zv}\right)
\times
\nonumber\\
& &
\exp\left(\frac{(-|Y|^2 + |P|^2 - 2i Y_1 P_1 - 2i Y_2 P_2)z}{1-zv}\right)
\nonumber\\
&=&
\frac{1}{\pi^2} \frac{1}{1-zv}
\exp\left(-\frac{(|Y|^2 + |P|^2)(1+zv)}{1-zv}\right)
\times
\nonumber\\
& &
\sum_{i=0}^{\infty} \frac{(-|Y|^2 + |P|^2 + 2i Y_1 P_1 + 2i Y_2 P_2)^i v^i}{i!(1-zv)^i}
\times
\nonumber\\
& &
\sum_{j=0}^{\infty} \frac{(-|Y|^2 + |P|^2 - 2i Y_1 P_1 - 2i Y_2 P_2)^j z^j}
{j!(1-zv)^j}~.
\end{eqnarray}
We have calculated the finite form of the series
$\sum_{m, n = 0}^{\infty} W_{mn} z^m v^n$, but what we really want
in order to get $W_{NN}$\ is the finite form of the series
$\sum_{m=0}^{\infty} W_{mm} u^m$.
This means that we must select terms of equal order in $z$\ and $v$ in the 
double sum and replace the product $zv$\ by $u$.
Looking at the last expression
(\ref{finform}), we see that the terms that must be kept are those
with $i=j$\ in the two sums. We can perform the transformation
and write the relevant generating functional:
\begin{eqnarray}\label{gfw}
\sum_{m=0}^{\infty} W_{mm} u^m
& = &
\frac{1}{\pi^2} \frac{1}{1-u} \exp\left(-\frac{(|Y|^2 + |P|^2)(1+u)}{1-u}
\right)\times\nonumber\\
& &
\sum_{i=0}^{\infty} \frac{{\left( (|Y|^2 - |P|^2)^2 + (2 Y_1 P_1 +2 Y_2 
P_2)^2 \right) }^i u^i}{(i!)^2{(1-u)}^{2i}}
\nonumber\\
& = &
\frac{1}{\pi^2} \frac{1}{1-u} \exp\left(-\frac{(|Y|^2 + |P|^2)(1+u)}{1-u}
\right)\times\nonumber\\
& & \!\!
J_0 \! \left(\! \frac
{2 {\left[ -u \left( {(|Y|^2 - |P|^2)}^2 + {(2 Y_1 P_1 +2 Y_2 P_2)}^2 
\right) \right]}^{1/2}}{1-u}\right)~,
\end{eqnarray}
where $J_0$ is the Bessel function.

It is now possible to identify the generating functional of a product
of Laguerre polynomials by using the identity (eq. 8.976.1 in~\cite{GR}):
\begin{equation}
\sum_{m=0}^{\infty} L_m(x) L_m(y) u^m =
\frac{1}{1-u} \exp \left( -\frac{(x+y)u}{1-u} \right)
J_0 \left( \frac{2{(-xyu)}^{1/2}}{1-u}\right)\label{gensqrlag}~.
\end{equation}
If we take:
\begin{eqnarray}
x & = & |Y|^2 + |P|^2 + 2 Y_1 P_2 - 2 Y_2 P_1  =  {|Y-iP|}^2~, \nonumber \\
y & = & |Y|^2 + |P|^2 - 2 Y_1 P_2 + 2 Y_2 P_1  =  {|Y+iP|}^2~,
\end{eqnarray}
we can finally write (\ref{gfw}) as:
\begin{equation}
\sum_{m=0}^{\infty} W_{mm} u^m =
\frac{1}{\pi^2} e^{-(|Y|^2 + |P|^2)} \sum_{m=0}^{\infty}
L_m(x)L_m(y) u^m~.
\end{equation}
This gives the final result:
\begin{eqnarray}
W_{mm} & = &
\frac{1}{\pi^2} e^{-(|Y|^2 + |P|^2)}~L_m( {|Y-iP|}^2 )~L_m( {|Y+iP|}^2 )
\nonumber \\
& = & W_{00}~L_m( {|Y-iP|}^2 )~L_m( {|Y+iP|}^2 )~.
\end{eqnarray}
Transition to semiclassical behaviour can be analyzed using this result. We
see from the last expression that the Wigner function, again in the limit of
infinite squeezing $|r_k|\to \infty$, will be delta-concentrated around 
classical trajectories
of the system. Indeed, $W_{00}$, the Wigner function corresponding to the 
initial vacuum satisfies~\cite{cqg96}
\begin{equation}
W_{00}\to |\Psi_0|^2~\delta~(P)
\end{equation}
and, therefore, also
\begin{equation}
W_{NN}\to |\Psi_N|^2~\delta~(P) \label{phdist}
\end{equation}
where $\delta (P)$ is the $\delta$-distribution. Thus, we obtain one more
independent derivation of the relation (\ref{trans}) between $y$ and $p$
in the limit $|r_k|\to \infty$ which results in the reduction of (\ref{equiv}) 
to (\ref{heis}).

An interesting point is that, in contrast to the vacuum
initial state where the Wigner function is positive definite, here the Wigner 
function is no longer positive definite {\em before} the limiting transition
$|r_k|\to \infty$ leading to (\ref{phdist}) is performed. Thus, $W$ cannot be 
interpreted as a probability
distribution in classical phase space for any finite $|r_k|$. This property 
actually constitutes the core of the criticism of the Wigner function approach
in the papers \cite{anderson90}. However, 
we do not want to insist on this classical interpretation of the Wigner 
function before the limiting transition (why should we?), and we use the 
Wigner function only to show that, for high squeezing, it 
becomes concentrated near a bunch of 
classical trajectories in phase space with initial stochastic amplitude
$y(\bf{k},\eta_o)$ obeying the probability density function (\ref{distr}).
For this reason, too, we deliberately use the word ``concentrated'' instead 
of ``peaked'' because the latter term, though often used may create a wrong 
impression about the positivity of the fine structure of the Wigner function 
around these classical trajectories. On the other hand, {\em after} the 
limiting transition $|r_k|\to \infty$ the Wigner function acquires the 
structure (\ref{phdist}) which {\em may} be used as a classical probability 
distribution in phase space. Moreover, this probability distribution is 
sufficient for all problems where oscillating quantum fine structure 
(``wave pattern'') around classical trajectories can be neglected.

This discussion also sheds some new light on the question as to what extent 
any coarse graining is necessary to obtain the transition to semiclassical 
behaviour of quantum cosmological perturbations. A coarse graining becomes
really necessary if one intends to obtain some classical probability
distribution in phase space from the Wigner function before the limiting
transition $|r_k|\to \infty$. However, for quantities for which the above 
mentioned quantum 
fine structure is unimportant (and all quantities observable in forseeable
future belong to this class) all parameters resulting from the correct
(i.e. physically justified) coarse graining disappear after the transition 
$|r_k|\to \infty$, so we are allowed to forget about coarse graining at all. 
As a result, for these quantities, decoherence (understood as a partial 
loss of quantum coherence, with the remaining part of coherence being 
expressable 
in classical terms) is reached without consideration of any actual
mechanism of coarse graining and decoherence (that is why we call it
``decoherence without decoherence''). On the other hand, there do exist
quantities and problems for which this fine structure is important (e.g., it
is essential for the calculation of the entropy of cosmological 
perturbations). But then one should not introduce a coarse graining simply
by hand or average over some degrees of freedom of the Universe arbitrarily
called as ``unobservable''. All such choices have to be justified by 
consideration of real physical processes.

The very specific case of a coherent initial state, for which the
Wigner function is given by eq. (\ref{coh}), is still instructive since it
clearly shows a smooth transition between deterministic and purely
stochastic behaviour of perturbations in the semiclassical limit.
Indeed, if $|\langle y(\eta) \rangle|^2\gg \max \lbrace {1\over 2k},
~|f_k(\eta)|^2\rbrace$, 
we get an effective deterministic classical behaviour (a time-dependent Bose
condensate of the field $y$). Realization of this inequality at all times 
is possible, e.g., by choosing initial conditions $\langle y \rangle_0$
and $\langle p\rangle_0$ proportional to $f_k(\eta_0)$ and $g_k(\eta_0)$
resp. with a large coefficient of proportionality. In the opposite limit
$|f_k(\eta)|^2\gg \max \lbrace|\langle y(\eta) \rangle|^2,~{1\over 2k}
\rbrace$ characteristic for
real cosmological perturbations, the average value of perturbations is
negligible with respect to their {\it rms} value. Then quantum perturbations
become equivalent (in the sense defined above) to classical quantities with 
a stochastic amplitude, a fixed phase defined by the $\delta$ multipliers in 
the expression for the Wigner function (\ref{coh}), and a practically zero 
average value. Of course, generally both average and {\it rms} values may be
non-zero. We see that the appearance of deterministic quantities
in the semiclassical limit is a very specific phenomenon restricted to
a narrow class of both special initial states and special time evolution
(with not too many ``particles''created). In particular, even the choice of a
coherent initial state which is already classical and deterministic
at the initial moment $\eta_0$ does not guarantee that it remains so at later 
times. A constructive counterexample is provided by the case where the
initial conditions $\langle y \rangle_0$ and $\langle p\rangle_0$ correspond
to the pure decaying mode in the long-wavelenghth regime $\lambda H\gg 1$.
On the contrary, transition to effective stochastic classical behaviour
is a generic effect which arises as a result of high squeezing of a very
wide class of initial quantum states. 
   
\section{Thermal state}

Until now we have considered N-particles initial states. Let us now consider 
the important case of a thermal initial state. Such a state arises for 
example in the
cosmological scenario where the stage of ``thermal inflation'' \cite{lyth95} 
takes place after relaxation of the radiation-dominated Universe to the state
of thermal equilibrium if we mean by
$\eta_0$ the moment when this stage of late inflation begins.
In this case, the probability density in the $y({\bf k})$-representation
is given by :

\begin{eqnarray}
{\cal P} \left( y({\bf k}),y(-{\bf k}) \right)
& = &
(1-e^{-\frac{1}{N_0}})~\langle y | \left( \sum_{N=0}^{\infty} e^{-N/N_0}
| N \rangle \langle N | \right) | y \rangle \nonumber\\
& = &
A(N_0)~\sum_{N=0}^{\infty} e^{-N/N_0}
\psi_N^{\ast} \left(  y({\bf k}) \right)
\psi_N \left(  y({\bf k}) \right)~.
\end{eqnarray}
where $N_0$ is the ratio of the temperature to the energy field quanta for 
given ${\bf k}$ and $A(N_0)$ is a normalization constant.
So, actually $N_0$ depends on $k$ but we drop the $k$ for 
brevity of notation. The probability to find $N$ quanta in the initial state
drops rapidly for  $N>N_0$.

This probability density can be calculated using the generating functionnal
of squared Laguerre polynomials, already given in (\ref{gensqrlag}):
\begin{eqnarray}
{\cal P} \left(|y|) \right)
& = &
\frac{(1-e^{-\frac{1}{N_0}})}{\pi~|f_k|^2}\left\{ \sum_{N=0}^{\infty}
e^{-N/N_0}~L_N^2 \left( \frac{|y|^2}{|f_ k|^2} \right) \right\}
\exp\left(- \frac{|y|^2}{|f_k|^2} \right) \nonumber\\
& = &
\left\{ \exp\left(- \frac{|y|^2}{|f_k|^2}
\frac{2e^{-1/A}}{1-e^{-1/N_0}} \right)
J_0\left( \frac{|y|^2}{|f_k|^2}
\frac{2ie^{-1/2N_0}}{1-e^{-1/N_0}} \right) \right\} \nonumber\\
& \times &
\frac{1}{\pi~|f_k|^2}\exp\left(- \frac{|y|^2}{|f_k|^2} \right) \nonumber\\
& = &
\frac{1}{\pi~|f_k|^2} \exp\left(- \frac{|y|^2}{|f_k|^2}\coth \frac{1}{2N_0}
\right) J_0\left( \frac{|y|^2}{|f_k|^2}\frac{i}{\sinh \frac{1}{2N_0}} \right)~.
\end{eqnarray}

When $N_0\ll1$, {\it i.e.}\ the temperature is much smaller than the energy
of the quanta for given ${\bf k}$ ,
we see that ${\cal P}$ is almost Gaussian,
since the exponential dominates the Bessel function. This is of course
to be expected since then our thermal state almost reduces to the vacuum
state, there are practically no thermal excitations. What can be said if 
$N_0$ grows, from what temperature on will we get a noticeable departure 
from a Gaussian probability density ?
It turns out that for any value of $N_0$, the probability ${\cal P}$ is a
symmetric single-peaked
function centered around zero, at first sight quite similar to a gaussian.
Therefore, the relevant discriminating parameter to be calculated in order
to study the gaussianity of this distribution is the kurtosis.
Fortunately, the latter can be found analytically.
Since ${\cal P}$\ depends only on $|y|$, its even
momenta are given by the following integrals:

\begin{eqnarray}
I_{2n}
& = & \int_{0}^{\infty} |y|^{2n} {\cal P}(|y|) 2\pi |y| d|y| \nonumber\\
& = & |f_k|^{2n} \int_{0}^{\infty} x^n \exp(-x \coth \frac{1}{2N_0})~
J_0 \left( \frac{ix}{\sinh \frac{1}{2N_0}} \right) dx~.
\end{eqnarray}
Four different integrals of the Bessel functions are given in~\cite{GR}, 
6.621.1, with the results expressed in terms of the hypergeometric function
$F$. Using the third one, we check the normalisation $I_0=1$.
With the second one, we calculate $I_2$\ and $I_4$, and finally the kurtosis
$Q$:

\begin{eqnarray}
&I_2&
= |f_k|^2~\coth \frac{1}{2N_0} \; F \left( -\frac{1}{2},0;1;\cosh^{-2}
\frac{1}{2N_0} \right) = |f_k|^2 \coth \frac{1}{2N_0}~, \\
&I_4&
= 2 |f_k|^4~\coth^2 \frac{1}{2N_0} \; F \left( -1,-\frac{1}{2};1;
\cosh^{-2} \frac{1}{2N_0} \right)
= |f_k|^4 \frac{1 + 2 \cosh^2 \frac{1}{2N_0}}{\sinh^2 \frac{1}{2N_0}}~, \\
&Q&
= I_4 / (I_2)^2 - 2 = \cosh^{-2} \frac{1}{2N_0}~.\label{kurt}
\end{eqnarray}

\begin{figure}[htbp]
\epsfysize=10cm
$$
\epsfbox{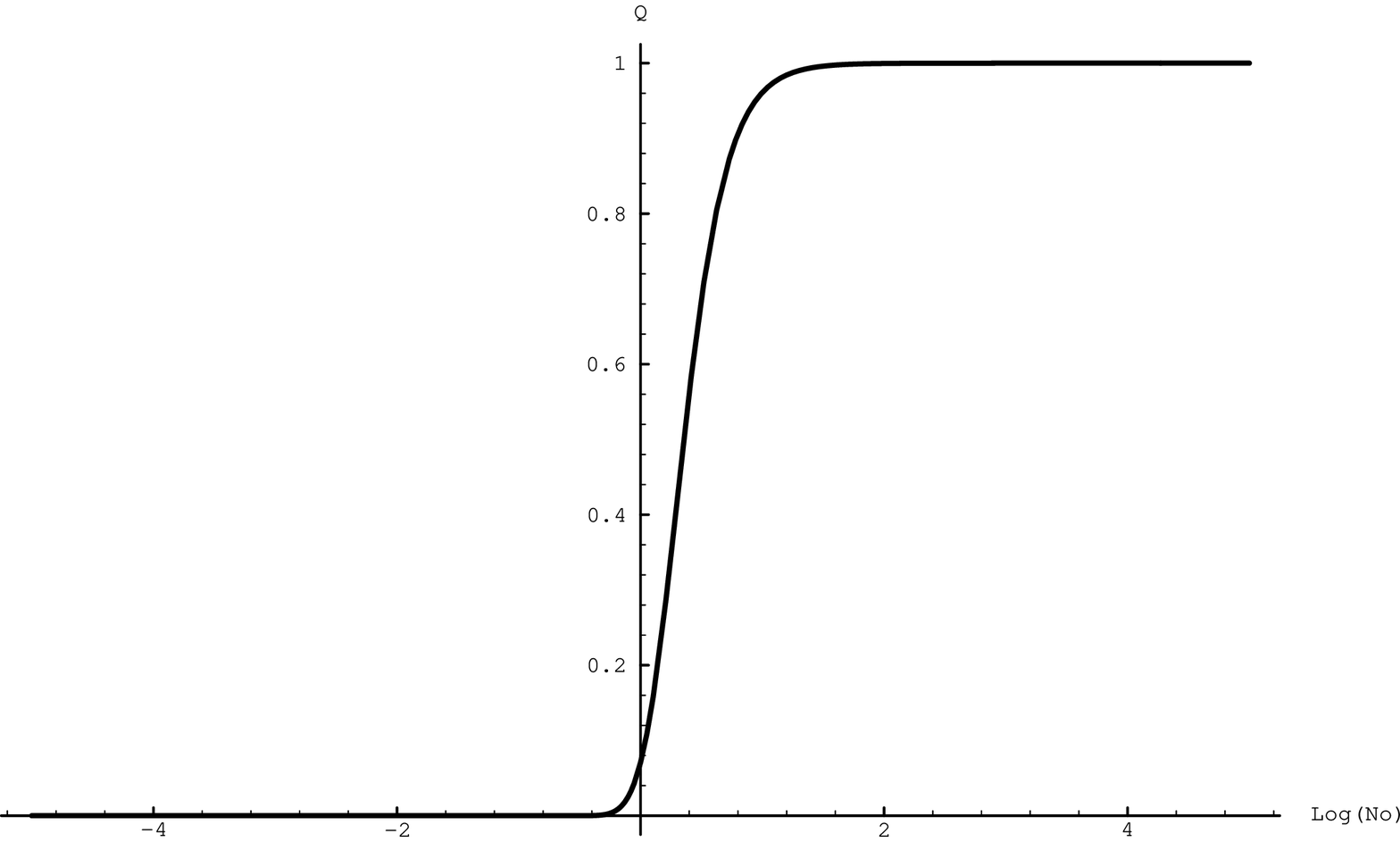}
$$
\caption[]{The kurtosis $Q$ as a function of $\log (N_0)$ for a squeezed 
thermal state specified by $N_0$ (for given ${\bf k}$).} 

\label{kurtosis}
\end{figure}

\begin{figure}[htbp]
\epsfxsize=10cm
$$
\epsfbox{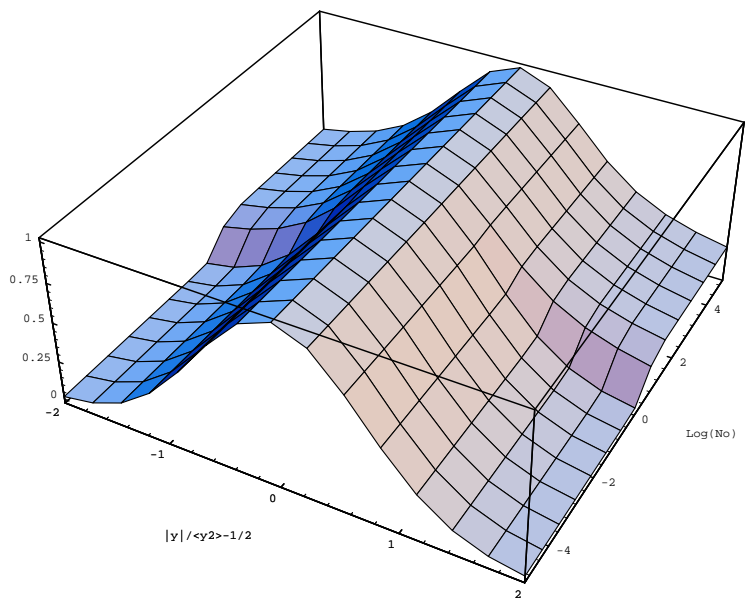}
$$
\caption[]{The probability distribution as a function of $\log (N_0)$ 
again for a squeezed thermal state. 
In order
to visualize the shape of the distribution and not its width, we plot
${\cal P}$, for given $N_0$, as a function of $|y|/\sqrt{<y~y^{\dagger}>}$.
We see that for $N_0\ll 1$, it is basically Gaussian.}
\label{probability}
\end{figure}

Looking at fig.\ref{kurtosis}, we see that there is a sharp transition between
a low-temperature gaussian behaviour ($Q = 0$) and a high-temperature
non-gaussian behaviour ($Q = 1$). Intermediate values of the kurtosis
imply that $N_0$ is close to one, within only one order of magnitude.
Fig.\ref{probability} shows how the gaussian transforms into another
distribution at high temperature. Scales of cosmological interest were
inside, most of them even deep inside, the Hubble radius at the onset of
inflation. Their energy is therefore much higher than the temperature at that
time. This corresponds to values of $N_0\ll 1$ for which, as can be seen from
~(\ref{kurt}) and from fig.\ref{probability}, the probability looks very
much like a Gaussian.

\section{Conclusions and discussion}

We have shown here that, for a wide class of initial non-vacuum states 
including all $N$-particles, thermal and most of the coherent states, 
quantum inhomogeneous cosmological 
perturbations (both adiabatic perturbations and gravitational waves) 
generically acquire semiclassical behaviour in the regime when their 
wavelenghth is much larger than the Hubble radius. In the language of quantum 
field theory, the latter regime corresponds to the high squeezing limit
$|r_k|\to \infty$. Then an equivalent classical description of these
perturbations in terms of c-number stochastic quantities becomes possible.
This, according to our paradigm, constitutes the transition from 
quantum to effectively classical behaviour of cosmological perturbations.
Thus, we generalize the results of~\cite{cqg96} obtained for the perturbations
of the vacuum initial state. The main difference between vacuum and 
non-vacuum initial conditions is that the statistics of the stochastic
amplitude of the equivalent classical field modes is Gaussian in the former 
case but generally non-Gaussian in the latter case. 

We proved this equivalence in three different independent ways. 
First, we explicitly constructed the wave
function in the Schr\"odinger representation thereby exhibiting its WKB
shape. Then, we have shown that the operators in the Heisenberg
representation correspond to classical stochastic functions in that limit.
Finally, we have computed the Wigner function and shown that it becomes
concentrated along classical trajectories. Since we do not in general
choose a unique vacuum state from an infinite number of unitary 
non-equivalent vacuum states at the initial moment $\eta_0$ (the case
of the inflationary scenario where a unique choice is possible was used 
as an illustration only), our results are
valid for $N$-particles, thermal and coherent states constructed using
any possible vacuum state. From a mathematical point of view, this ambiguity
is reflected in the freedom to choose the constant parameters $C_1({\bf k})$ 
and
$C_2({\bf k})$ characterizing the long-wavelength regime (\ref{A1}) which
satisfy only one normalization condition (\ref{A2}).

It turns out that the generalization from vacuum to non-vacuum initial 
conditions
does not produce any new problem of principle for the quantum-to-classical 
transition.
The only thing which is important for it is that the classicality conditions 
for perturbations derived in section 4 should be satisfied (in particular,
it is enough to have a sufficienly large amplitude of the
quasi-isotropic mode of perturbations in the long-wavelength regime).
The Wigner function for $N$-particles or thermal initial states (but not for 
coherent
initial states) is not positive definite for a finite $r_k$. However, we have
shown that this does not prevent the transition to the semiclassical
behaviour in the limit $|r_k| \to \infty$ expressed by eq. (\ref{phdist}).
We believe that this addresses the criticism against the transition to
the semiclassical behaviour raised in~\cite{anderson90}
since the generalization of our considerations to the case where the scale 
factor $a(\eta)$ is itself quantized but still in the WKB regime does not
bring new principally different features.

Note also that we never use the so called ``semiclassical gravity'', or 
one-loop approximation, $R_{ik}=8\pi G\langle T_{ik}\rangle$ to study the 
evolution of quantum perturbations and the quantum-to-classical transition
for them (these equations are usually understood as equations for the average 
value $\langle g_{ik}\rangle$ of the metric tensor). This approximation is 
inappropriate for the problem under consideration, not only 
because one-loop approximation is not exact when only a few quantum fields
are important, but mainly due to the fact that these equations are grossly 
insufficient in the typical case arising in cosmology where average values
of perturbations $\langle \delta g_{ik}\rangle$ are negligible (or even 
exactly zero) as compared to their {\it rms} values, and all their
higher moments should be calculated (see also the discussion at the end of 
section 5). So, one should go beyond the one-loop, or the Gaussian, 
approximation.

Now we discuss the question whether this non-Gaussianity of perturbaions
which appears in the case of non-vacuum initial states is observable, and if
it can be used to exclude such states using observational data on ${\Delta T
\over T}$ angular fluctuations of the CMB temperature. First, one should  
take into account that observable quantities in physical space are real and
consist of sums of quantities of the type ($y({\bf k, \eta})e^{i{\bf kr}}+
~c.c.$), irrespective of the fact whether $y$ is an operator or a c-number 
stochastic quantity. As a result, the statistics of quantities in real space 
is the same as
for one-mode states. In particular, in the Gaussian case the probability 
distribution is the one-dimensional Gaussian distribution, not the
two-dimensional one that prevails for the complex amplitude $y({\bf k},
\eta)$. Then the value of the curtosis $Q$ for quantities in real space in the
case of an $N$-particles initial state is given by the expression 
(\ref{onemode})(e.g., $\varphi$ can be taken as ${\Delta T\over T}$ itself).
Previously, the skewness $S\equiv \langle \left({{\Delta T}\over T}\right)^3
\rangle$ and topological characteristics, like the genus, of the 
CMB temperature
fluctuations were mainly considered. However, $S=0$ for all $N$-particles
initial states. So, we propose to measure the kurtosis $Q$ since this is
more interesting. For non-vacuum states which we considered, $Q$ is negative.  

For a non-Gaussian distribution, it is easy to see that $Q$ lies below the
cosmic variance level of purely Gaussian perturbations for all initial 
non-vacuuum
states where different ${\bf k}$ modes are $\delta$-correlated and {\it rms} 
values of 
perturbations depend on $|{\bf k}|$ only. This results due to the averaging
over angular directions which is equivalent to summing over all ${\bf k}$
modes with the same $|{\bf k}|$. Such a sum will rapidly become Gaussian 
due to the central limit theorem. In particular, the curtosis of the sum of 
$n$ equally distributed modes with kurtosis $Q$ is equal to $Q/n\sim 
1/n$ while the cosmic variance limit for the Gaussian distribution 
$\sim 1/\sqrt n$. Still, the non-zero kurtosis can be observable for less
symmetric and more correlated initial states. An extreme example is when
${\Delta T(\theta,\phi)\over T}$ can take only 2 values $\pm \Delta_0$
with equal probability. Then the kurtosis $Q=-2$ and it certainly lies above 
the cosmic variance level. On the other hand, this discussion shows that
there is a wide class of initial non-vacuum states which cannot be
distinguished from the vacuum initial state by looking at the statistics
of observational quantities, so it is not easy to disprove them.

Finally, we have shown the crucial role played by the quasi-isotropic mode
of perturbations in the quantum-to-classical transition. Namely, if this
mode is present then, as time goes on and the Universe expands, perturbations 
become more and more semiclassical. On the contrary, this does not take 
place for the decaying mode. Note that there is no quasi-isotropic mode in the
case of the exactly homogeneous and isotropic FRW model (even fully 
quantized) or of the cosmological model of the Bianchi I type. There are
known problems with the quantum-to-classical transtion in FRW quantum
cosmology (even for large values of the scale factor) summarized in 
paper~\cite{claus95}. We will not discuss here to what extent these problems
are really dangerous for the FRW cosmology, but we would like to point out that
the results of our paper suggest that the final solution to these problems 
will require consideration of more general, inhomogeneous cosmological models
which contain the quasi-isotropic mode of perturbations (or its generalization
to the fully non-linear case). Also, it is tempting to connect 
the arrow of time in the Universe with the quantum-to-classical transition.
Then a possible relation between the arrow of time and the growth of 
inhomogeneities in the Universe (the transition from order to chaos) emerges.  

\vspace{1cm}
\noindent
{\bf Acknowledgements}
\par\noindent
A significant part of this project was accomplished during the visit of
one of the authors (A.S.) to France under
the agreement between the Landau Institute for Theoretical Physics and
Ecole Normale Sup\'erieure, Paris. A.S. thanks ENS and EP93 CNRS (Tours) for
financial support and Profs E. Brezin, C. Barrabes for their hospitality
in ENS, Universit\'e de Tours respectively. A.S. also acknowledges financial
support by the Russian Foundation for Basic Research, grant 96-02-17591.

\end{document}